\documentclass[12pt]{article}
\usepackage{amssymb,amsmath,epsfig,cite}

\numberwithin{equation}{section}
 \allowdisplaybreaks

\begin{document}
\title{\textbf{Thermodynamics and Glassy Phase Transition of Regular
Black Holes}}
\author{Wajiha Javed$^1$ \thanks{wajiha.javed@ue.edu.pk; wajihajaved84@yahoo.com},
Z. Yousaf$^2$
\thanks{zeeshan.math@pu.edu.pk} and Zunaira Akhtar$^1$ \thanks{zunaira35akhtar@gmail.com}\\
$^1$Division of Science and Technology,\\ University of Education,
Township Campus, Lahore, Pakistan\\
$^2$Department of Mathematics, University of the Punjab, \\Quaid-i-Azam Campus, Lahore-54590, Pakistan}
\date{}
\maketitle
\begin{abstract}
The present paper is aimed to study thermodynamical properties of
phase transition for regular charged black holes. In this context,
we have considered two different forms of black hole metrics
supplemented with exponential and logistic distribution functions
and investigated the recent expansion of phase transition through
grand canonical ensemble. After exploring the corresponding
Ehrenfest's equation, we found the second order background of phase
transition at critical points. In order to check the critical
behavior of regular black holes, we have evaluated some
corresponding explicit relations for the critical temperature,
pressure and volume and draw certain graphs with constant values of
Smarr's mass. We found that for the black hole metric with exponential
configuration function, the phase transition curves are divergent
near the critical points, while glassy phase transition has been
observed for the Ay\'{o}n-Beato-Garc\'{i}a-Bronnikov black hole in $n=5$
dimensions.
\end{abstract}

\section{Introduction}

In general relativity (GR), the black hole (BH) is defined as a region
of space having intense gravitational field so that no matter or no radiations can
escape from it. After entering into any BH, both matter and light entrapped as such stellar
structures are impregnated with spacetime singularity. Three physical
terminologies like momentum, charge and mass, could provide all the descriptions
of any BH structures. Depending upon the presence of these variables,
various terminologies have been given to various types of BHs, such as
Kerr BH, Schwarzschild BH, Reissner-Nordstr\"{o}m BH etc. Hawking attempted to resolve a paradox
between gravitational theories and quantum mechanics and declared that the claim that the light can not escape
within the horizon, may not be true. Recently, a great amount of work has been done in investigating the BH
solutions including the background of higher dimensions \cite{zee1,zeen1,zeen2}. Rodriguez \cite{zee2} was the among the pioneers
who classified higher dimensional BH species. It is well-known that
the corrections beyond the environment of semiclassical approximation
could give rise to corrected form of entropy for Myers-Perry BHs and Hawking temperature.

The study of BH thermodynamics gives interrelationship among the laws of thermodynamics and BH mechanics. There are two intuitive routes to BH thermodynamics, i.e., the laws of classical thermodynamics and BH dynamics. In GR, BHs obey certain laws that have some mathematical correspondence with the ordinary laws of thermodynamics. In classical theory, they are perfect absorbers and does not emit anything,
thereby providing absolute zero temperature. However, in the realm od quantum theory, they emit Hawking radiations with
a perfect thermal spectrum. The presence of BH and quantum effects lead to the violation of second law of
thermodynamics and area theorem. Initially, when Bekenstein \cite{Z3,Z4} proposed generalized version of second law of thermodynamics,
he did not consider the possibility of the decrease in area.

About more than fifty years ago, Stephen Hawking \cite{zee3} claimed that BH is not fully black rather
such systems are involved in emitting radiations at a particular value of
temperature given by $T=\frac{\kappa}{2\pi}$, where $\kappa$ is the horizon's surface gravity. This
temperature formulation is widely known as Hawking temperature. This discovery has open a new research window to explore various concealed aspects of unfounded quantum gravity. This has also raised multiple questions about the theoretical
mathematical analogies between laws BH mechanics and thermodynamics that were previously
elaborated by Bardeen et al. \cite{zee4}.
The attraction of researchers towards phase transition in BHs is an example in this context. Phase transition
and its consequences has been started by the seminal work of Davis
\cite{zee5}. He found that such a phenomenon could only happened, if Kerr-Newman BHs reach
at the point of divergent heat capacity. In this direction, a great amount of work has been done in investigating the
phase transition in various models of BHs.

Biswas and Chakraborty \cite{zee6} considered classical and topological choices of BH thermodynamics and checked the applicability of phase transition in H\u{o}rava Lifshitz gravity. Sharif and Wajiha \cite{Z26} observed Hawking radiation through tunneling aspiring from the charged regular
Bardeen and Ay\'{o}n-Beato-Garc\'{i}a-Bronnikov (ABGB) BHs and found that found the contribution
of charge and energy in the drawing of radiation spectrum. Gergely et al. \cite{zee5} discussed the role of thermodynamical variables, like heat capacity and phase transition on the evolving chatged Bhs and no phase transition has been observed after performing
Poincar\'{e} stability analysis. Poshteh et al. \cite{zee7}
explored even dimensional phase transition of Myers-Perry BH and evaluated the corresponding Ehrenfest's equations
by assuming equal components of angular momenta. They noticed second-order phase transition
in the light of semiclassical regime. Dolan et al. \cite{zee8} explored mathematical correspondence between thermodynamic and geometric volumes for the Kerr-Newman de Sitter BH in four dimensional space. They concluded that Chong-Cvetic-Lu-Pope solution
could possibly demonstrate BH solution with gravitational corrections coming from
Einstein-Chern-Simons theory. Mansoori and Mirza \cite{zee9} explored the problem of relating thermodynamical properties of
various types of BHs and found a mathematical expression between curvature singularities and heat capacity. Yousaf
et al. \cite{zeem1, zeem2, zeem3} found relationships between matter variables and geometrical coefficients of the celestial relativistic bodies and checked the role of extra curvature corrections in their dynamics. Abbas et al. \cite{ab1} found some interesting results based on the
dynamical evolution of black holes and stellar interiors.

Zou et al. \cite{zee10} discussed thermodynamical properties of
Born-Infeld anti de Sitter BHs with $D$-dimensional background. They
found the existence of phase transition in an environment of four or
above small BH, which is analogous to results that obtained with the
Van de Waals liquid gas. Tharanath et al. \cite{zee11} investigated
some thermodynamical features of a regular Bardeen BH and confronted
with second order thermodynamic phase transitions equation.

Pourhassan et al. \cite{zee12} studied the influences of thermal
variations on the relativistic modified Hayward BH and first law of
thermodynamics and found stable BH configurations due to the
presence of thermal fluctuations. Miao et al. \cite{zee13} examined
the role of noncommutative and original thermodynamical pressure
components in the realm of negative $\Lambda$-dominated epoch and
concluded that such pressure components tend to produce opposite
effects in the BH phase transitions. Chaturvedi et al. \cite{zee14}
analyzed the phenomenon of phase transitions for $4$ dimensional
charged anti de Sitter BH in order to explore some interesting
results. They found divergent behavior of the thermodynamical scalar
invariant at some point which is compatible with analysis of van der
Waals gas. Taranath and Suresh \cite{Z24} have splendid work on
phase transition and geometrothermodynamics of regular BHs and its
singularities in heat capacity. Saleh et al. \cite{Z23} discussed
phase transition of Bardeen regular BH and
thermodynamics.Furthermore, examine about reveal phase transition of
general regular BH (distribution function) \cite{Z2} as well as
logistic distribution function \cite{Z21}. It is established that
the second order thermal phase transition occurs in ordinary BHs.
Balart and Vagenas \cite{Z1} presented various formulations of
charged regular BH spacetimes supplemented with special matter
content functions motivated from the background of continuous
probability distributions. Further, we discuss Smarr's mass by
considering first law of BH thermodynamics, which is usually holds
for regular BHs in non-linear electrodynamics \cite{Z17}

The paper is organized as under. In the next section, we provide different quantitative as well as qualitative
aspects of phase transition for our regular black holes. We have also investigated the nature of charged BH for
inspired with exponential distribution variables. Section \textbf{3} is devoted to perform
the same analysis for the case of ABGB BH under some modification\cite{Z18}. We have also various graphs in order to investigate
various thermodynamical features of phase transition. Conclusion and discussion are mentioned in the last section.

\section{Charged Regular BH in Higher Dimensions}

The static form of a spherical BH can be represented with the help of the following spacetime \cite{Z26}
\begin{equation}\label{2.1}
ds^{2}=-f(r)dt^{2}+f(r)^{-1}dr^{2}+r^{2}(d\theta^{2}+sin^{2}\theta d\phi^{2}),
\end{equation}
where the metric coefficient $f(r)$ is given by \cite{Z1}
\begin{eqnarray}\label{2.2}
f(r)&=&1-\frac{2M}{r}\exp\left(-\frac{q^{2}}{2Mr}\right),
\end{eqnarray}
in which $M$ and $q$ correspond to ADM mass and charge occupied by the BH.
The function $f$ disappears at the location of horizons. In order to locate
outer horizon (OH) and inner horizon (IH), we must use $f(r_{\pm})=0$, the
corresponding OH and IH are found.
\begin{align}\label{2.3}
r_{+}=-\frac{q^{2}}{2MW(0,\frac{-q^{2}}{4M^{2}})},\quad
r_{-}=-\frac{q^{2}}{2MW(-1,\frac{-q^{2}}{4M^{2}})},
\end{align}
where subscripts $-$ and $+$ stand for IH and OH, respectively. In the above expression,
the quantity $W$ specifies Lambert's $W$ function. The quantities associated with the
surface gravities ($\kappa$) at IH and OH are found, respectively, as
\begin{eqnarray}\label{2.4}
\kappa_{+}=\frac{1}{2}\left.\frac{d f(r)}{dr}\right|_{r=r_{+}}
&=&\left(\frac{-m(r(\sigma(r_{+}))^{'}-(\sigma(r_{+})))}{r^{2}_{+}}\right),\\\label{2.5}
\kappa_{-}=\frac{1}{2}\left.\frac{d f(r)}{dr}\right|_{r=r_{-}}
&=&\left(\frac{-m(r(\sigma(r_{-}))^{'}-(\sigma(r_{_{-}})))}{r^{2}_{-}}\right).
\end{eqnarray}
\begin{figure}\center
\epsfig{file=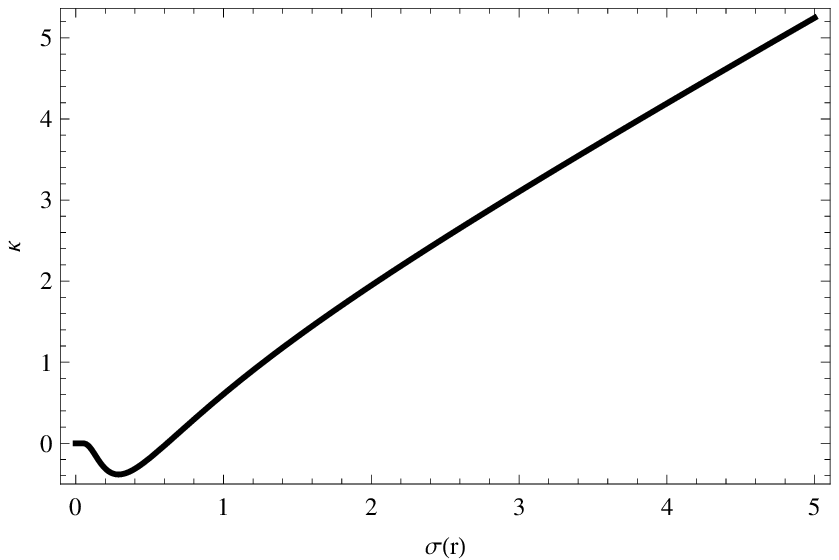,width=0.48\linewidth}\epsfig{file=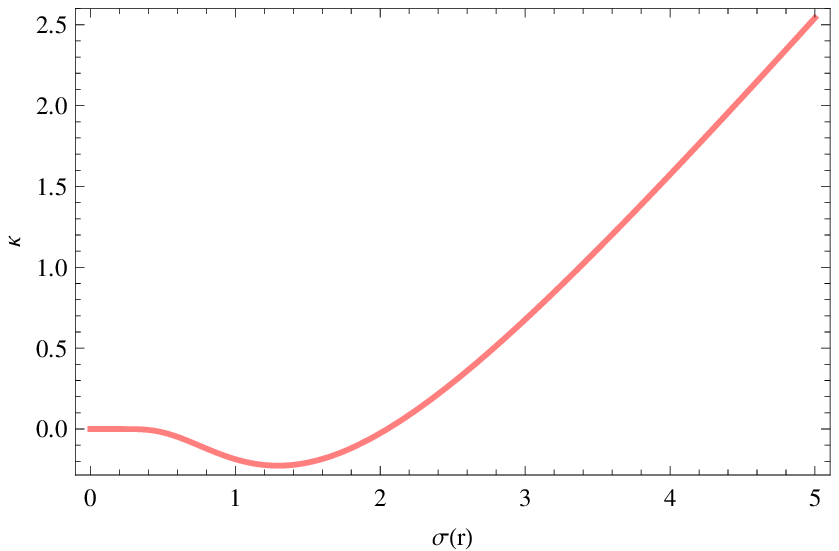,width=0.48\linewidth},
\epsfig{file=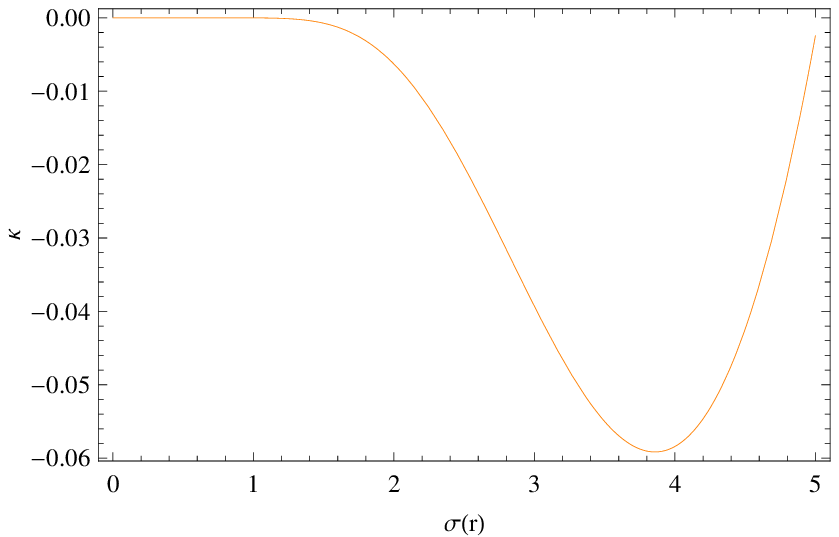,width=0.48\linewidth}\caption{Behavior of surface gravity $\kappa$ corresponding to $\sigma(r)$.}
\label{f1}
\end{figure}
The thermodynamical terms associated with OH and IH, after retaining
first thermodynamics law, yield
$(S_{+},~S_{-},~\phi_{+},~\phi_{-},~V_{+},~V_{-},~P=0$ (where
$\Lambda=0$)). For the sake of BH thermodynamics in Einstein gravity
coupled to non-linear matter, the first law of BH thermodynamics
corresponding to horizons can be expressed in the following form
\cite{Z17}
\begin{eqnarray}\label{2.6}
\delta m =\frac{\kappa_{+}}{2\pi}\delta S_{+}+\phi_{+}\delta q+V_{+}\delta P,\\\label{2.7}
\delta m =\frac{\kappa_{-}}{2\pi}\delta S_{-}+\phi_{-}\delta
q+V_{-}\delta P,
\end{eqnarray}
where $m$ is Smarr's mass \cite{Z2}, in our analysis of phase
transition it can be considered as a constant.

The above equations can also be expressed as
\begin{eqnarray}\label{2.8}
d S_{+}=\frac{2\pi}{\kappa_{+}}(\delta m-\phi_{+}\delta q-V_{+}\delta P),\\\label{2.9}
d S_{-}=\frac{2\pi}{\kappa_{-}}(\delta m-\phi_{-}\delta
q-V_{-}\delta P).
\end{eqnarray}
So far, we have evaluated mathematical terms related with thermodynamics without invoking
the correspondence of OH and IH. It has been observed that there are two variables in the
charged regular BH, i.e., $m$ and $q$, therefore thermodynamical variables associated with
OH and IH must be the functions of $m$ and $q$. The thermodynamical
entropy ($S$) with the background of charged regular spacetime
can be calculated by adding the OH and IH, therefore we have
\begin{eqnarray}\nonumber
S=S_{+}+S_{-},
\end{eqnarray}
which has an alternative form as
\begin{equation}\label{2.10}
dS=dS_{+}+dS_{-}.
\end{equation}
After taking zero value of cosmological constant along with Eqs.(\ref{2.6}) and (\ref{2.7}), Eq.(\ref{2.10}) can be recasted
\begin{align}\label{2.11}
dS=2\pi\left({\frac{1}{\kappa_{+}}}+{\frac{1}{\kappa_{-}}}\right)d m
-\frac{nq}{2(n-1)}\left({\frac{1}{r_{+}^{n-1}\kappa_{+}}}+
\frac{1}{r_{-}^{n-1}\kappa_{-}}\right)dq-2\pi\frac{Vol(S^{n})}{(n+1)}
\left(\frac{r_{+}^{n-1}}{\kappa_{+}}+\frac{r_{-}^{n-1}}{\kappa_{-}}\right)dP.
\end{align}
By making use of $P=0$ as well as $\Lambda=0$ \cite{zee8}, Eq.(\ref{2.11}) becomes
\begin{equation}\label {1.9}
dS=2\pi\left({\frac{1}{\kappa_{+}}}+{\frac{1}{\kappa_{-}}}\right)dm-
\frac{nq}{2(n-1)}\left({\frac{1}{r_{+}^{n-1}\kappa_{+}}}+\frac{1}{r_{-}^{n-1}\kappa_{-}}\right)dq.
\end{equation}
The probability configurations corresponding to Eq.(\ref{2.1}) is represented with the help of
the following exponential function
\begin{equation}\nonumber
\sigma=\frac{1}{\exp(x)}.
\end{equation}
This function obey the constraint $\sigma(x)>0$ along with $d\sigma/dx<0$ with $x\geq0.$ Equations
(\ref{2.4}) and (\ref{2.5}) give rise to
\begin{eqnarray}\label {2.13}
dr_{+}=\frac{1}{2\kappa_{+}}\left(-mr_{+}^{2}(\sigma(r_{+}))^{''}\right),\\\label {2.14}
dr_{-}=\frac{1}{2\kappa_{-}}\left(-mr_{-}^{2}(\sigma(r_{-}))^{''}\right).
\end{eqnarray}
In the formulations of thermodynamics, the volume of higher-dimensional regular BH
was defined, in generally, as \cite{Z2}
\begin{eqnarray}\nonumber
V=\left(\frac{\partial m}{\partial P}\right)_{S^{n},q,\sigma(r)...},
\end{eqnarray}
which gives rise to
\begin{eqnarray}\label{2.15}
V=\frac{VolS^{n}}{n+1}(r_{-}^{n+1}-r_{+}^{n-1}), \quad
dV=Vol(S^{n})\left(r_{-}^{n}d r_{-}-r_{+}^{n}d r_{+}\right).
\end{eqnarray}
By making use of Eqs.(\ref{2.13}) and (\ref{2.14}) in Eq.(\ref{2.15}), we get
\begin{equation}\label{2.16}
dV=Vol(S^{n})\left[\frac{1}{2\kappa_{-}}\left[m
r_{-}^{2}(\sigma(r_{-}))^{''}\right]r_{-}^{n}-\frac{1}{2\kappa_{+}}
\left[m r_{+}^{2}(\sigma(r_{+}))^{''}\right]r_{+}^{n}\right],
\end{equation}
which after using the value of $\sigma(r)$ turns out to be
\begin{eqnarray}\label{2.17}
dV=\frac{m}{2}Vol(S^{n})[\frac{r_{-}^{n+2}}{\kappa_{-}}(-\frac{\exp\frac{-q^{2}}{2mr_{-}}}{mr_{-}}+
(\frac{\exp\frac{-q^{2}}{2mr_{-}q^{2}}}{m^{2}r_{-}^{2}}))\nonumber\\
-(\frac{r_{+}^{n+2}}{\kappa_{+}}(-\frac{\exp\frac{-q^{2}}{2mr_{+}}}{mr_{+}}+
(\frac{\exp\frac{-q^{2}}{2mr_{+}q^{2}}}{m^{2}r_{+}^{2}})))]dq.
\end{eqnarray}
Equation (\ref{2.17}) after some manipulation provides
\begin{align}\label{2.18}
dq=\frac{dV}{\frac{m}{2}Vol(S^{n})\left[\frac{r_{-}^{n+2}}{\kappa_{-}}
\left(-\frac{\exp(\frac{-q^{2}}{2mr_{-}})}{mr_{-}}+
\frac{\exp(\frac{-q^{2}}{2mr_{-}q^{2}})}{m^{2}r_{-}^{2}}\right)-
\left(\frac{r_{+}^{n+2}}{\kappa_{+}}(-\frac{\exp(\frac{-q^{2}}{2mr_{+}})}{mr_{+}}+
\frac{\exp(\frac{-q^{2}}{2mr_{+}q^{2}})}{m^{2}r_{+}^{2}})\right)\right]}.
\end{align}

\section{Phase Transition in the Spacetime of Charged ABGB BH}

The analysis of the phase transition of various BH models
has achieved great attraction among relativistic astrophysicists.
The special similarity relations between asymptotical BH models and the
Van der Waals expressions have been observed by many researchers
in literature. [16-19,23-26]. The exploration of phase transition for BHs experiencing non-equilibrium
phases of non equilibrium system is very interesting.
The thermodynamical variables respecting IH must need to associated with
that of OH. In order to study thermodynamical
features of the charged regular BH metric in an appropriate way, one may need to
assume relationships between the OH and IH horizons.
In this paper, we wish to discuss effects related to phase transition of
the regular charge BH. We wish to draw various figures in order to highlight the behaviors
of thermodynamical quantities like entropy and temperature. It is well-known that
Eherfest's equation describes the nature of phase transition and now
we shall consider the Van der Waal's equation of the form
\begin{eqnarray}\label{3.1}
P=\frac{kT}{\nu-b}-\frac{a}{\nu^{2}},
\end{eqnarray}
where $P,~T,~k$ indicate pressure, temperature and Boltzmann constant, respectively. Furthermore,
the terms $\nu\equiv V/N,a~,b$ demonstrate specific values of fluid volume and their parameters, respectively.
This equation could be used to examine $P-\nu$ plots with some fixed $T$ values. Upon substituting the values of
critical points and conditions, one can evaluate expressions for
critical temperature, pressure and volume.

\begin{figure}\center
\epsfig{file=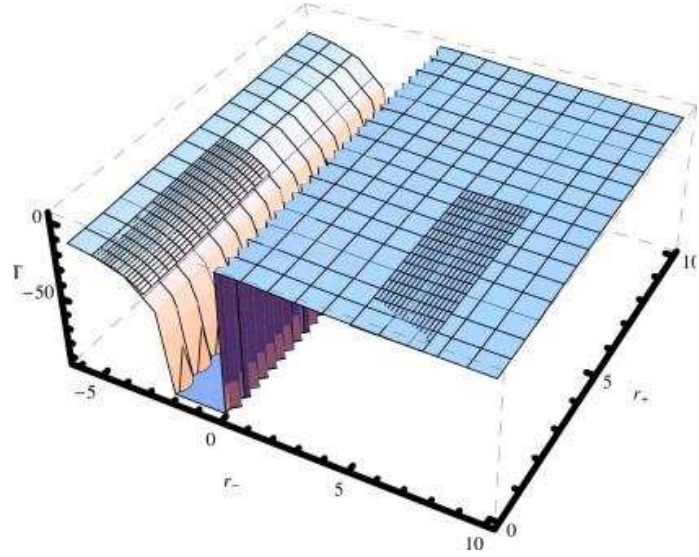,width=0.55\linewidth}
\caption{Consequence of temperature ($T_{\sigma(r)}$) when outer and inner horizons coincides.}\label{f2}
\end{figure}

\begin{figure}\center
\epsfig{file=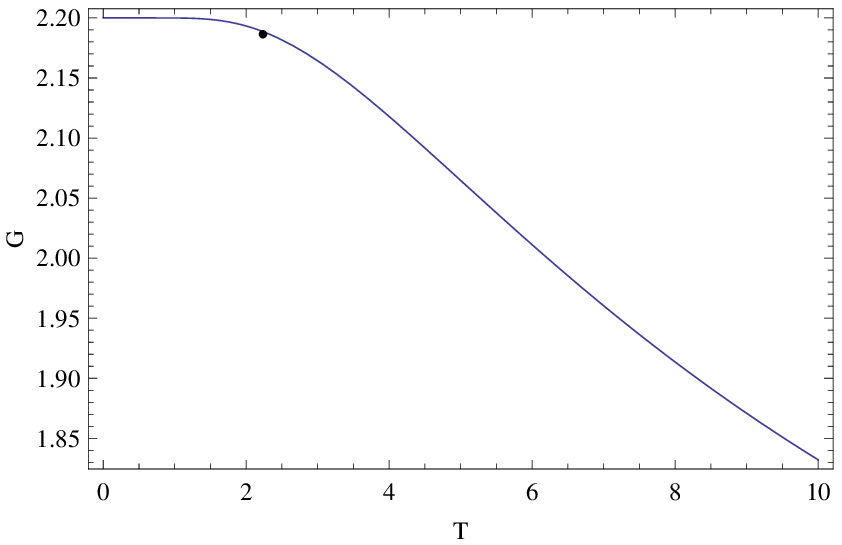,width=0.4\linewidth}
\epsfig{file=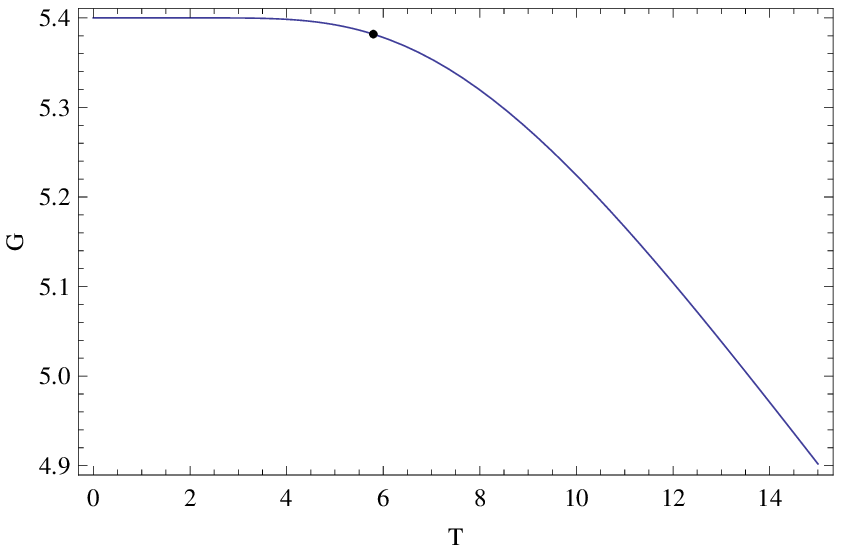,width=0.4\linewidth}\epsfig{file=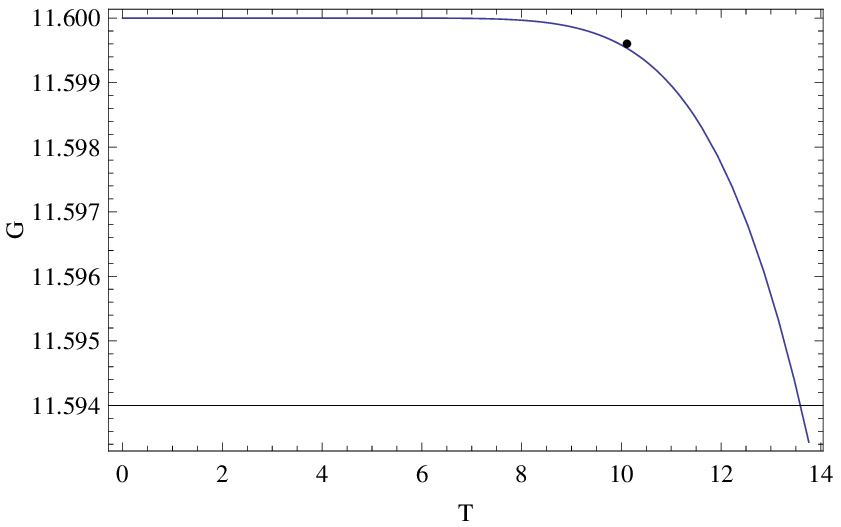,width=0.4\linewidth}\caption{Illustration of Gibbs
 free energy for $q=2,~q=5,~q=11$.}\label{f3}
\end{figure}

\begin{figure}\center
\epsfig{file=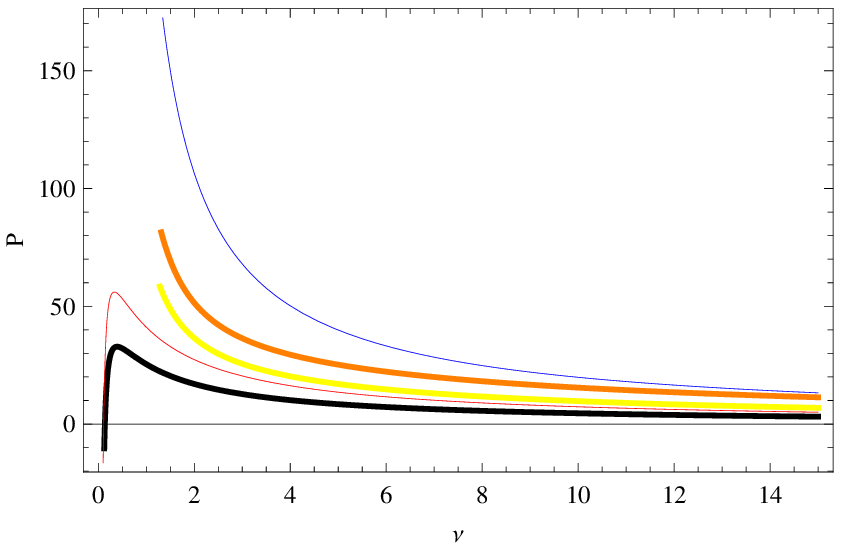,width=0.6\linewidth}\caption{$P-\nu$ curves for fixed values of $q$.}\label{f4}
\end{figure}

The behavior of surface gravity with respect of $\sigma$ for our regular BH has been
mentioned in Fig.\ref{f1}. We have checked the consequences of temperature under the consition when OH and IH coincides. This has
been shown in Fig.\ref{f2}. Figure \ref{f3} illustrates the fluctuations of Gibbs free energy for
$q=2,~q=5$ and $q=11$. One can observe from these graphs about the
variations of Gibbs energy $(G)$ with different values of entropy.
It is interestingly noticed that plots are moving downwards by
increasing entropy and vice versa. To check the correspondence of $P$ and $\nu$ in the neighborhood of
critical values of temperature, we have drawn various curves of $P-\nu$
with different temperature backgrounds. Figure \ref{f4} describes that
the stable condition $\left(\frac{\partial P}{\partial \nu}\right)_{T_{eff}}<0$ can be achieved by setting $T_{eff}>T^c_{eff}$.
It can be seen that on setting the little value of $\nu$, the quantity
$\left(\frac{\partial P}{\partial \nu}\right)_{T_{eff}}>0$ thereby indicating the unstable
phase of the relativistic system and opposite results can be observed for large $\nu$
values. This suggested the possible occurrence of phase transition at $T_{eff}=T_{eff}^c$.
Moreover, we also plotted some curves of
$S-T,~C-P^c$ and $\zeta-\nu$ as shown in Figs.\ref{f5}, \ref{f6} and \ref{f7}, respectively.
From these plots, one can observe the effects of thermodynamics quantities.
We found that the quantities $S$ and $G$ are behaving continuously at the corresponding critical points. The
same behavior has been observed in the case of Reissner-Nordstr\"{o}m
metric by Zhang \emph{et al.} \cite{Z2}.

One can find the modified equation for the first law of thermodynamics in Ref. \cite{Z19}
and for the formulation of critical points $(\phi,q=constant)$ then equation becomes
\begin{eqnarray}\label{3.2}
dm=TdS+VdP,
\end{eqnarray}
where
\begin{align}\nonumber
T&=2\pi\rho_{o},\\\nonumber
P&=\left(\frac{r_{+}^{-n}-r_{+}^{-n}\rho_{2}}{\rho_{o}}+\frac{\rho_{2}(\rho_{1})}{\rho_{o}}\right),
\end{align}
for
\begin{align}\nonumber
\rho_{o}&=[-m(r_{+}(\sigma(r_{+})^{'})-\sigma(r_{+})]-[m(r_{-}(\sigma(r_{-})^{'})-\sigma(r_{-}))],\\\nonumber
\rho_{1}&=\left[-\frac{\sigma(r_{+})}{mr_{+}}+\frac{\sigma(r_{+})q^{2}}{m^{2}r_{+}^{2}}\right]+\left[\frac{\sigma(r_{-})}{mr_{-}}+
\frac{\sigma(r_{-})q^{2}}{m^{2}r_{-}^{2}}\right],\\\nonumber
\rho_{2}&=\left[r_{+}^{3}-r_{-}^{3}\right].
\end{align}
By putting the values of $\kappa_{+}$ and $\kappa_{-}$ from Eqs.(\ref{2.4}) and (\ref{2.5}), we get
\begin{align}\nonumber
dm&=2\pi\left[\frac{r_{+}^{2}}{-m(r_{+}(\sigma(r_{+})^{'})-\sigma(r_{+}))}
-\frac{r_{-}^{2}}{m(r_{-}(\sigma(r_{-})^{'})-\sigma(r_{-}))}\right]^{-1}d S+2\pi\\\nonumber
&\times\left[\frac{r_{+}^{2}}{-m\left(r_{+}(\sigma(r_{+})^{'})-\sigma(r_{+})\right)}
-\frac{r_{-}^{2}}{m\left(r_{-}(\sigma(r_{-})^{'})-\sigma(r_{-})
\right)}\right]^{-1}\frac{nq}{m(n-1)}\\\nonumber
&\times(\frac{r_{+}^{3-n}}{-m\left(r_{+}(\sigma(r_{+})^{'})-\sigma(r_{+})\right)}
-\frac{r_{-}^{3-n}}{-m\left(r_{-}(\sigma(r_{-})^{'})-\sigma(r_{-})\right)}
\times\frac{r_{-}^{3}}{-m\left(r_{-}(\sigma)^{'}-\sigma(r_{-})\right)}\\\label{3.3}
&\left(-\frac{\sigma(r_{-})}{mr_{-}}+\frac{\sigma(r_{-})q^{2}}{m^{2}r_{-}^{2}}\right)
-\frac{r_{+}^{3}}{-m\left(r_{+}(\sigma)^{'}-\sigma(r_{+})\right)}
\left(-\frac{\sigma(r_{+})}{mr_{+}}+\frac{\sigma(r_{+})q^{2}}{m^{2}r_{+}^{2}}\right))dV.
\end{align}
From some interesting speculation $\gamma=\kappa_{+}+\kappa_{-}$, we have
\begin{eqnarray}\label{3.4}
dm=2\pi\gamma d S+\left[2\pi\gamma+r^{3-n}\gamma^{}-1+r^{3}(-\frac{\sigma(r_{+})}{mr_{+}}
+\frac{\sigma(r_{+})q^{2}}{m^{2}r_{+}^{2}})\gamma^{-1}\right]dV.
\end{eqnarray}
Furthermore, we study the intersecting property of thermodynamics (heat capacity), 
stability and instability of any BH represented by heat capacity, which
is negative to produce Hawking radiation and BH will be unstable because phase
transition does not exist due to negative heat capacity but it can be positive if BH
adduces charge at the critical point. structure of liquid{gas does not change 
unanticipated in Van der Waals system second-order the phase transition which will be
occurred when the second derivative of potential under distribution function is 
continuous and then we can check the behavior of higher-dimensional (ABGB) toward
the critical points.

The specific heat of the regular charged BH with non-varying pressure $P_{\sigma(r)}$,
charge $q$ and volume thermal expansion
coefficient $\zeta$ can be evaluated as
\begin{align}\nonumber
C_{P(\sigma(r))}&=T_{c}\left(\frac{\partial S}{\partial T_{c}}\right)_{P(\sigma(r))}=\frac{\pi(r_{-}^{2}-r_{+}^{2})}{2}\left\{\frac{4m}{(r_{-}(\sigma(r_{-}^{'})))}\right\}
\left\{\frac{\partial(p(\sigma(r)))}{\partial(\zeta)}\right\}q\\\nonumber
&-\left[\left(\frac{nq^{2}r_{-}^{-n}\log(m)}{(-1+n)(-1+r_{-})(\sigma
r_{-})^{2}}\right)-\left(\frac{nq^{2}r_{+}^{-n}\log(m)}{(-1+n)(-1+r_{+})(\sigma
r_{+})^{2}}\right)\right]\\\label{3.5}
&\times\left[2\pi\left(-mr_{-}(\sigma(r_{-}^{'}))-
\sigma(r_{-})\right)-\left(-mr_{+}(\sigma(r_{+}^{'}))-\sigma(r_{+})\right)\right],\\\label{3.6}
\zeta&=\frac{1}{\nu}\frac{\nabla V}{\nu_{o}}\left[2\pi-m(r_{c}\frac{\exp\frac{-q^{2}}{2mr_{c}}}{mr_{c}}
+\frac{\exp\frac{-q^{2}}{2mr_{c}}q^{2}}{m^{2}r_{c}^{2}}-
\frac{\exp\frac{-q^{2}}{2mr_{c}}}{mr_{c}})\right].
\end{align}
\begin{figure}\center
\epsfig{file=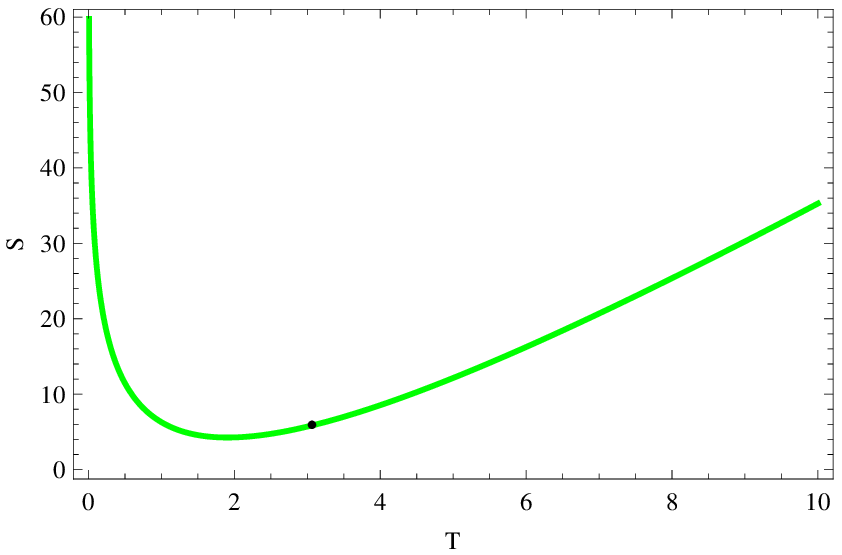,width=0.48\linewidth}\epsfig{file=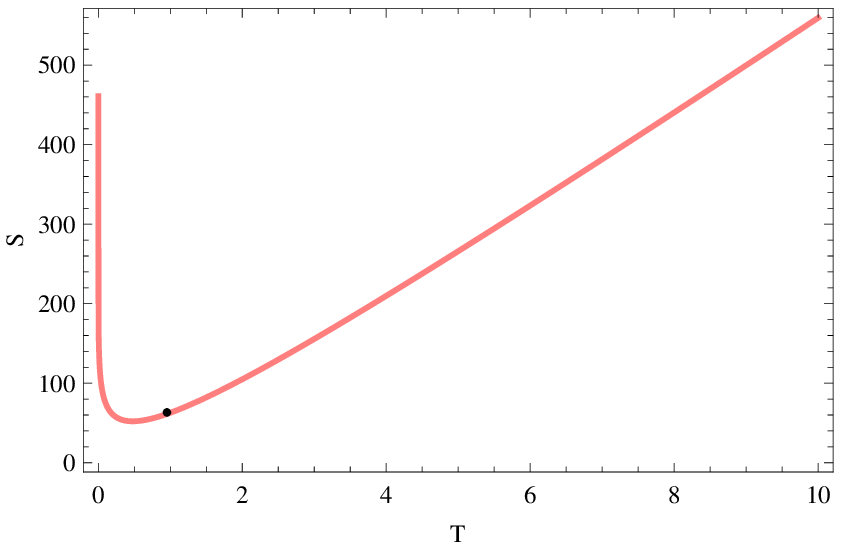,width=0.48\linewidth},
\epsfig{file=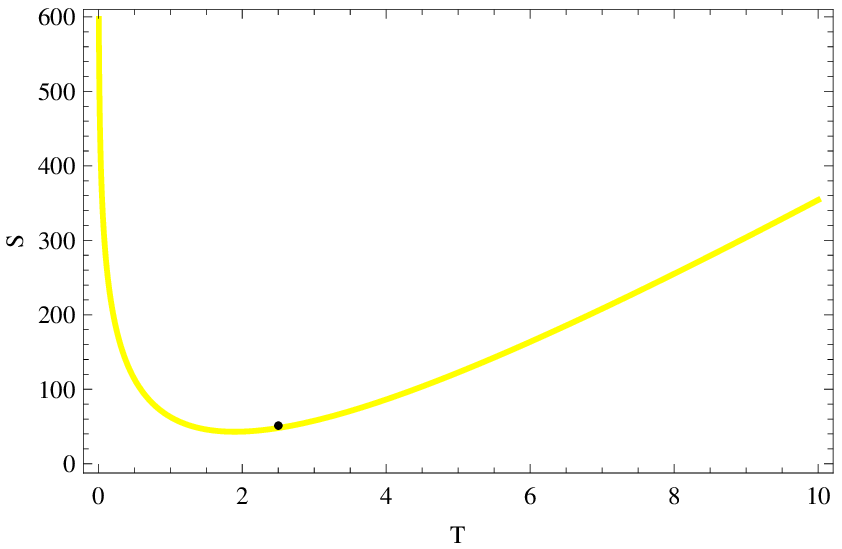,width=0.48\linewidth}\caption{$S_{\sigma(r)}-T_{\sigma(r)}$ curves for a charged regular BH.}\label{f5}
\end{figure}
\begin{figure}\center
\epsfig{file=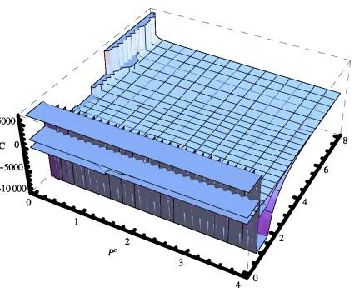,width=0.55\linewidth}\caption{Heat capacity corresponding to $P^{c}_{\sigma(r)}$ .}\label{f6}
\end{figure}
Moreover, we plot the curves of $\kappa-\sigma(r)$, $\zeta-\nu$,
$3D$ plot of $T_{\sigma(r)}-r_{+}r_{-}$ and $C-P^{c}_{\sigma(r)}$ in
Figs. 1, 2, 6 and 7, respectively, which show that the surface gravity
corresponding to distribution function, behavior of temperature
corresponding to horizons and last one is heat capacity.
\begin{figure}\center
\epsfig{file=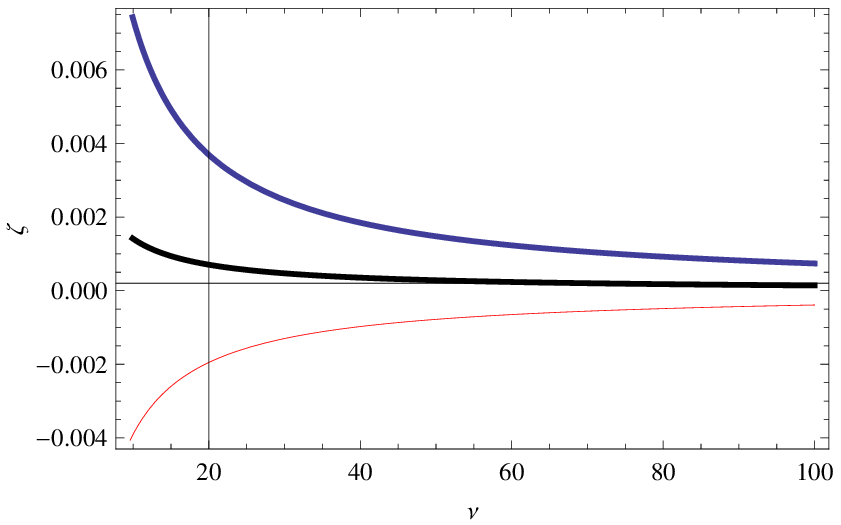,width=0.6\linewidth}\caption{Coefficient of volume thermal expansion $\zeta$ verses $\nu$.}\label{f7}
\end{figure}
we consider $P_{\sigma(r)}=P$ and $T_{\sigma(r)}=T$. When $n=5$ for
$q$ is invariant Now we find out the phase transition for splendid
cases,and calculate the critical points for different cases $q=2, 5,
11$ respectively in space time dimension $n=5$ by these cases
position of critical horizon is decreases as well as critical
temperature, critical pressure, but the critical specific Mass,
entropy, Gibbs free energy constant, volume thermal expansion
constant increases,then distinguish the relation of critical
pressure $P^{c}_{\sigma(r)}$ and volume $V^{c}_{\sigma(r)}$ by
$p-\nu$ $[Fig:4]$  and $(T^{c}_{\sigma(r)})$ is also involve in
specific stipulation. for stability of system we stimulate the
General condition by exponential form (contemplate approximation)
(1.2) also using equation (3.9) we create Table $1$ of numerical
solution for
$(T^{c}_{\sigma(r)}),(P^{c}_{\sigma(r)}),(V^{c}_{\sigma(r)})
,(G^{c}_{\sigma(r)}),(S^{c}_{\sigma(r)}),(\zeta^{c}_{\sigma(r)}),
(r^{c}_{\sigma(r)})$ for $q=2, 5, 11$ in $n=5$ dimension
\begin{equation}\label{3.7}
\left(\frac{\partial P}{\partial v_{\sigma(r)}}\right)_{T}=0,\left(\frac{\partial^{2} P}{\partial v^{2}_{\sigma(r)} }\right)_{T}=0.
\end{equation}
\begin{table}[ht]
\centering 
\begin{tabular}{c c c c} 
\hline\hline 
cases & $q=2$ & $q=5$ & $q=11$ \\ [0.5ex] 
\hline 
$T$ & 0.05241 & 0.05123 & 0.0312 \\ 
$P$ & 0.1562 & 0.0542 & 0.2318 \\
$m$ & 15.872 & 24.514 & 18.427 \\
$S$ & 9.456 & 88.341 & 97.898 \\
$r$ & 2.7851 & 1.4521 & 2.4741 \\
$G$ & 2.1801 & 5.3021 & 11.4329 \\
$\zeta$ &0.0025 & 0.0075 & -0.0043 \\ [1ex ] 
\hline 
\end{tabular}
\caption{Numerical solutions of phase transition for $n=5$}
\label{table:critical points} 
\end{table}
We have Ehrenfest's equations for higher dimensional regular charged BH
in which subscript $1$ and $2$ personate the phase
transitions 1 and 2, respectively. Taking into account the analogy between
ABGB BH and the thermodynamical phase parameters, the Ehrenfest's equations can be recasted as
\begin{align}\label{3.8}
&-\left(\frac{\partial P}{\partial T_{\sigma(r)}}\right)_{q}=\frac{\zeta_{2}-\zeta_{1}} {\kappa_{2}-\kappa_{1}}=\frac{\Delta\zeta}{\Delta\kappa},\\\label{3.9}
&-\left(\frac{\partial P}{\partial T_{\sigma(r)}}\right)_{S}=\frac{C_{T_2}-C_{T_1}}{T_{\sigma(r)}\nu^c(\zeta_{2}-\zeta_{1})}
=\frac{\Delta C_{T}}{T\nu^c\Delta \zeta},
\end{align}
where $\kappa=\frac{1}{\nu}\left(\frac{\partial \nu}{\partial P}\right)_T$ and
$\zeta=\frac{1}{\nu}\left(\frac{\partial \nu}{\partial T}\right)_P$ are analogies
for the coefficient of volume expansion and isothermal compressibility, respectively.
These right hand side of the above equations are found to be finite at critical points.
In the realm of thermodynamics, according to our condition the Maxwell's equations
can be formulated as
\begin{equation}\label{3.10}
\left(\frac{\partial \nu}{\partial S}\right)_{T}=\left(\frac{\partial T}{\partial P}\right)_S.
\end{equation}
By making use of Eqs.(\ref{3.8}) and (\ref{3.10}), it follows that
\begin{equation}\label{3.11}
\left(\frac{\triangle\zeta}{\triangle\kappa}_{T}\right)=\left(\frac{\partial\nu }{\partial S}\right)_{T_{\sigma(r)}}.
\end{equation}
After evaluating some useful expressions and some mathematical manipulations, we get
\begin{equation}\label{3.12}
\left(\frac{\partial S}{\partial \nu }\right)^{c}_{T_{\sigma(r)}}=\left(\frac{\partial S}{\partial\nu}\right)^{c}_{P_{\sigma(r)}},
\end{equation}
thereby showing that Ehrenfesr expressions of our systems at the critical point are
true. Using above equation, the Prigogine-Defay ratio, $\prod$, can be expressed as
\begin{equation}\label{3.13}
\prod=\frac{\triangle C_{\sigma(r)}\triangle\kappa_{T}}{T_{\sigma(r)^{c}}\nu^{c}(\triangle\zeta)^{2}}=1,
\end{equation}
which shows that second order phase transition occurs at $T=T_0$.

\section{Glassy Phase Transition of Charged ABGB BH Using Logistic Distribution}

In the previous section, we have obtained phase transition of regular BH (distribution
function) \cite{Z1} and observed intersecting behavior of Gibbs free
energy, surface gravity, critical temperature, critical pressure and
volume thermal expansion coefficient $\zeta$ as well as
Bekenstein-Hawking entropy corresponding to IH and OH. Now, we wish to examine
the phase transition for ABGB BH supplemented with metric
function ($X(r)$ within same line element) given by (Logistic distribution
function)\cite{Z1}
\begin{align}\label{4.1}
X(r)=Y(r)=1-\frac{2M}{r}\left[\frac{4e^{-\sqrt\frac{2q^{2}}{\beta
Mr}}}{\left(e^{-\sqrt\frac{2q^{2}}{\beta
Mr}}+1\right)^{2}}\right]^{\beta},
\end{align}
where $\beta$ is a constant. The numerical values of horizons can be calculated by fixing $\beta$. The
metric coefficient of the well-known Schwarzschild BH can be obtained on setting $\beta\rightarrow0$. However,
the constraint $\beta\rightarrow\infty$, reduces the above metric variable with the following
\begin{align}\label{4.1}
X(r)=Y(r)=1-\frac{2M}{r}e^{\left(-\frac{q^{2}}{2
Mr}\right)}.
\end{align}
In this context, the first law of
thermodynamics and Bekenstein-Hawking entropy provide
\begin{align}\nonumber
&dr=\frac{r}{2\kappa}-2^{1+2\beta}\left(\frac{e^{-\sqrt2\sqrt\frac{q^{2}}{Mr\beta}}}
{(1+e^{-\sqrt2\sqrt\frac{q^{2}}{M r\beta}})^{2}}\right)^{\beta-1}
Mr\sqrt2e^{-2\sqrt2\sqrt\frac
{q^{2}}{Mr\beta}}q^{3}\left[\left\{\left(1+e^{-\sqrt2\sqrt\frac{q^{2}}{Mr\beta}}\right)^{3}\right.\right.\\\nonumber
&\left.\left.\times M^{2}r^{3}
\left(\sqrt\frac{q^{2}}{Mr\beta}\right)^{\frac{3}{2}}\beta^{2}\right\}^{-1}
-\frac{e^{-2\sqrt2\sqrt\frac {q^{2}}{Mr\beta}}q^{3}}
{\sqrt2\left(1+e^{-\sqrt2\sqrt\frac{q^{2}}{Mr\beta}}\right)^{3}M^{2}r^{3}
(\frac{q^{2}}{Mr\beta})}-6e^{-3\sqrt2\sqrt\frac{q^{2}}{Mr\beta}}q\right.\\\nonumber
&\left.\times\left(\left(1+e^{-\sqrt2\sqrt\frac{q^{2}}{Mr\beta}}\right)^{4}Mr^{2}\beta\right)^{-1}+
\frac{6e^{-2\sqrt2\sqrt\frac{q^{2}}{Mr\beta}}q}
{\left(1+e^{-\sqrt2\sqrt\frac{q^{2}}{Mr\beta}}\right)^{3}Mr^{2}\beta}
-\frac{e^{-\sqrt2\sqrt\frac{q^{2}}{Mr\beta}}}q\right.\\\nonumber
&\left.\times
{\left(1+e^{-\sqrt2\sqrt\frac{q^{2}}{Mr\beta}}\right)^{2}Mr^{2}\beta}
-\frac{2\sqrt2e^{-\sqrt2\sqrt\frac{q^{2}}{Mr\beta}}q}
{\left(1+e^{-\sqrt2\sqrt\frac{q^{2}}{Mr\beta}}\right)^{3}
Mr^{2}\beta\sqrt\frac{q^{2}}{Mr\beta}}\right.\\\nonumber
&\left.+\frac{\sqrt2e^{-\sqrt2\sqrt\frac{q^{2}}{Mr\beta}}q}
{\left(1+e^{-\sqrt2\sqrt\frac{q^{2}}{Mr\beta}}\right)^{2}Mr^{2}\beta\sqrt\frac{q^{2}}{Mr\beta}}\right]\beta
+2^{1+2\beta}(\frac{e^{-\sqrt2\sqrt\frac{q^{2}}{Mr\beta}}}
{\left(1+e^{-\sqrt2\sqrt\frac{q^{2}}{Mr\beta}}\right)^{2}})^{-1+\beta}\\\nonumber
&\times M\left(\frac{2\sqrt2e^{-2\sqrt2\sqrt\frac{q^{2}}{Mr\beta}}q}
{\left(1+e^{-\sqrt2\sqrt\frac{q^{2}}{Mr\beta}}\right)^{3}Mr\beta\sqrt\frac{q^{2}}{Mr\beta}}-
\frac{\sqrt2e^{-\sqrt2\sqrt\frac{q^{2}}{Mr\beta}}q}
{\left(1+e^{-\sqrt2\sqrt\frac{q^{2}}{Mr\beta}}\right)^{2}
Mr\beta\sqrt\frac{q^{2}}{Mr\beta}}\right)\beta\\\nonumber
&-2^{1+2\beta}(\frac{e^{-\sqrt2\sqrt\frac{q^{2}}{Mr\beta}}}
{\left(1+e^{-\sqrt2\sqrt\frac{q^{2}}{Mr^{2}\beta}}\right)^{2}})^{-2+\beta}Mr
\left\{-\frac{\sqrt2e^{-2\sqrt2\sqrt\frac{q^{2}}{Mr\beta}}q^{2}}
{\left(1+e^{-\sqrt2\sqrt\frac{q^{2}}{Mr\beta}}\right)^{3}Mr\beta\sqrt\frac{q^{2}}{Mr\beta}}\right.\\\nonumber
&\left.+\frac{e^{-\sqrt2\sqrt\frac{q^{2}}{Mr\beta}}q^{2}}{\sqrt2
\left(1+e^{-\sqrt2\sqrt\frac{q^{2}}{Mr\beta}}\right)^{2}Mr^{2}\beta\sqrt\frac{q^{2}}{Mr\beta}}\right\}
\left\{\frac{2\sqrt2e^{-2\sqrt2\sqrt\frac{q^{2}}{Mr\beta}}q}
{\left(1+e^{-\sqrt2\sqrt\frac{q^{2}}{Mr\beta}}\right)^{3}Mr\beta\sqrt\frac{q^{2}}{Mr\beta}}\right.
\\\label{4.3}
&\left.-\frac{\sqrt2e^{-\sqrt2\sqrt\frac{q^{2}}{Mr\beta}}q}
{\left(1+e^{-\sqrt2\sqrt\frac{q^{2}}{Mr\beta}}\right)^{2}Mr\beta\sqrt\frac{q^{2}}{Mr\beta}}\right\}(\beta-1)\beta.
\end{align}
For simplicity of formulation we plot the curves of
$\kappa-\beta$,$P-\beta$ and $G-\beta$ that represent the surface gravity,
critical pressure (including logistic function) \cite{Z21}
and free energy in grand canonical (Gibb's free energy)
ensembles at the critical point for $n=5$.
\begin{figure}\center
\epsfig{file=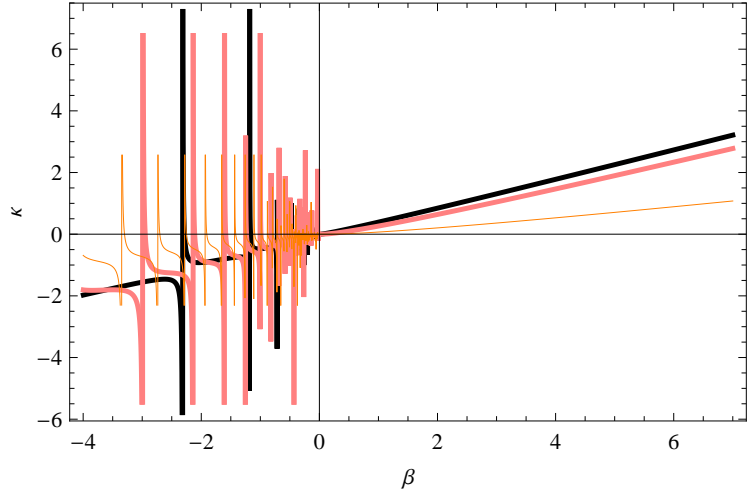,width=0.6\linewidth}\caption{surface gravity
corresponding to $\beta$.}\label{f8}
\end{figure}
\begin{figure}\center
\epsfig{file=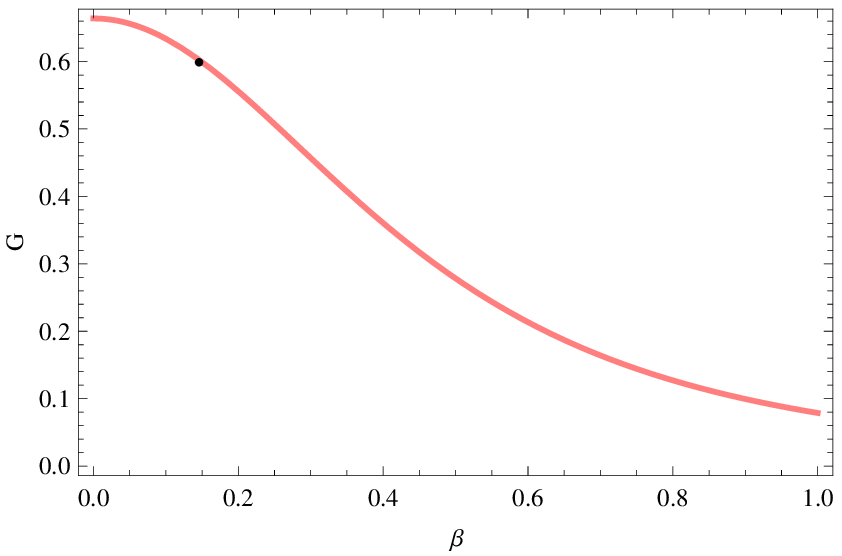,width=0.48\linewidth}\epsfig{file=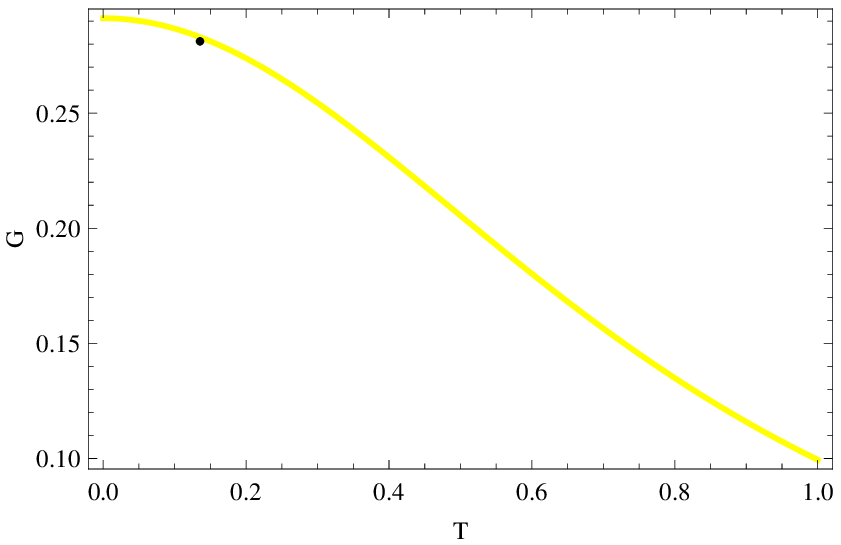,width=0.48\linewidth},
\epsfig{file=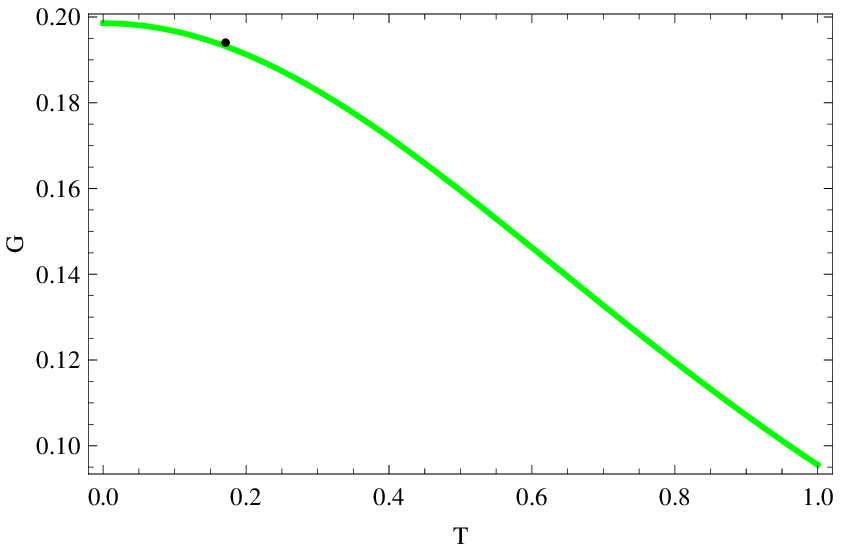,width=0.48\linewidth}\caption{$G-T$ curves for (ABGB) BH, Gibbs free energy $G=M-TS$ for $q=2,5,11$, respectively. Here we can observe behavior of second order derivative of Gibbs free energy corresponding to critical temperature at fixed charge $q$}\label{f9}
\end{figure}
With the background of thermodynamics, the volume of the higher dimensional space can be given as
\begin{align}\nonumber
&dV=\frac{Vol(S^{n})r^{-n}}{n+1}\left(\frac{1}{2\kappa}r\right)\left\{-2^{1+2\beta}
\left(\frac{e^{-\sqrt2\sqrt\frac{q^{2}}{Mr\beta}}}{\left(1+e^{-\sqrt2\sqrt
\frac{q^{2}}{Mr^{2}\beta}}\right)^{2}}\right)^{-1+\beta}Mr\right\}\\\nonumber&
\times\left[\frac{\sqrt2e^{-2\sqrt2\sqrt\frac{q^{2}}{Mr\beta}}q^{3}}
{\left(1+e^{-\sqrt2\sqrt\frac{q^{2}}{Mr\beta}}\right)^{3}M^{2}r^{3}\beta^{2}
\left(\frac{q^{2}}{Mr\beta}\right)^{\frac{3}{2}}}-
\frac{e^{-\sqrt2\sqrt\frac{q^{2}}{Mr\beta}}q^{3}}{\sqrt2
\left(1+e^{-\sqrt2\sqrt\frac{q^{2}}{Mr\beta}}\right)^{2}M^{2}r^{3}\beta^{2}}\right.\\\nonumber
&\left.\times \left(\frac{q^{2}}{Mr\beta}\right)^{-\frac{3}{2}}-
\frac{6e^{-3\sqrt2\frac{q^{2}}{Mr\beta}}q}{
\left(1+e^{-\sqrt2\sqrt\frac{q^{2}}{Mr\beta}}\right)^{4}Mr^{2}\beta}+
\frac{6e^{-2\sqrt2\frac{q^{2}}{Mr\beta}}q}{
\left(1+e^{-\sqrt2\sqrt\frac{q^{2}}{Mr\beta}}\right)^{3}Mr^{2}\beta}\right.\\\nonumber
&\left.- \frac{e^{-\sqrt2\sqrt\frac{q^{2}}{Mr\beta}}}q{
\left(1+e^{-\sqrt2\sqrt\frac{q^{2}}{Mr\beta}}\right)^{2}Mr^{2}\beta}
-\frac{2\sqrt2e^{-\sqrt2\sqrt\frac{q^{2}}{Mr\beta}}q}{
\left(1+e^{-\sqrt2\sqrt\frac{q^{2}}{Mr\beta}}\right)^{3}Mr^{2}\beta\sqrt\frac{q^{2}}{Mr\beta}}\right.\\\nonumber
&\left.+\frac{\sqrt2e^{-\sqrt2\sqrt\frac{q^{2}}{Mr\beta}}q}{
\left(1+e^{-\sqrt2\sqrt\frac{q^{2}}{Mr\beta}}\right)^{2}Mr^{2}\beta\sqrt\frac{q^{2}}{Mr\beta}}\right]\beta
+2^{1+2\beta}\left(\frac{e^{-\sqrt2\sqrt\frac{q^{2}}{Mr\beta}}}{
\left(1+e^{-\sqrt2\sqrt\frac{q^{2}}{Mr\beta}}\right)^{2}}\right)^{-1+\beta}\\\nonumber
&\times
M\left(\frac{2\sqrt2e^{-2\sqrt2\sqrt\frac{q^{2}}{Mr\beta}}q}{
\left(1+e^{-\sqrt2\sqrt\frac{q^{2}}{Mr\beta}}\right)^{3}Mr\beta\sqrt\frac{q^{2}}{Mr\beta}}-
\frac{\sqrt2e^{-\sqrt2\sqrt\frac{q^{2}}{Mr\beta}}q}{
\left(1+e^{-\sqrt2\sqrt\frac{q^{2}}{Mr\beta}}\right)^{2}Mr\beta\sqrt\frac{q^{2}}{Mr\beta}}\right)\beta\\\nonumber
&-2^{1+2\beta}\left(\frac{e^{-\sqrt2\sqrt\frac{q^{2}}{Mr\beta}}}{(1+e^{-\sqrt2\sqrt\frac{q^{2}}{Mr\beta}})^{2}}\right)^{-2+\beta}Mr
\left\{-\frac{\sqrt2e^{-2\sqrt2\sqrt\frac{q^{2}}{Mr\beta}}q^{2}}{
\left(1+e^{-\sqrt2\sqrt\frac{q^{2}}{Mr\beta}}\right)^{3}Mr\beta\sqrt\frac{q^{2}}{Mr\beta}}\right.\\\nonumber
&\left.+\frac{e^{-\sqrt2\sqrt\frac{q^{2}}{Mr\beta}}q^{2}}{\sqrt2
\left(1+e^{-\sqrt2\sqrt\frac{q^{2}}{Mr\beta}}\right)^{2}Mr^{2}\beta\sqrt\frac{q^{2}}{Mr\beta}}\right\}
-\frac{2\sqrt2e^{-2\sqrt2\sqrt\frac{q^{2}}{Mr\beta}}q}{
\left(1+e^{-\sqrt2\sqrt\frac{q^{2}}{Mr\beta}}\right)^{3}Mr\beta\sqrt\frac{q^{2}}{Mr\beta}}\\\label{4.4}
&-\frac{\sqrt2e^{-\sqrt2\sqrt\frac{q^{2}}{Mr\beta}}q}{
\left(1+e^{-\sqrt2\sqrt\frac{q^{2}}{Mr\beta}}\right)^{2}Mr\beta\sqrt\frac{q^{2}}{Mr\beta}}(-1+\beta)\beta,
\end{align}
which after substitution and manipulations gives
\begin{align}\nonumber
&A_{1}=\frac{2^{1+2\beta}\left(\frac{e^{-\sqrt2\sqrt\frac{q^{2}}{Mr\beta}}}{
\left(1+e^{-\sqrt2\sqrt\frac{q^{2}}{Mr\beta}}\right)^{2}}\right)^{\beta}M}{r^{2}}
-\frac{1}{r}2^{1+2\beta}\left(\frac{e^{-\sqrt2\sqrt\frac{q^{2}}{Mr\beta}}}{
\left(1+e^{-\sqrt2\sqrt\frac{q^{2}}{Mr\beta}}\right)^{2}}\right)^{-1+\beta}\\\nonumber
&\times
M\left(-\frac{\sqrt2e^{-2\sqrt2\sqrt\frac{q^{2}}{Mr\beta}}q^{2}}{(1+
e^{-\sqrt2\sqrt\frac{q^{2}}{Mr\beta}})^{3}Mr^{2}\beta\sqrt\frac{q^{2}}{Mr\beta}}+
\frac{e^{-\sqrt2\sqrt\frac{q^{2}}{Mr\beta}}q^{2}\left(\frac{q^{2}}{Mr\beta}\right)^{-1/2}}{\sqrt2
\left(1+e^{-\sqrt2\sqrt\frac{q^{2}}{Mr\beta}}\right)^{2}
Mr^{2}\beta}\right),\\\nonumber
&B=\left(12e^{-3\sqrt2\sqrt\frac{q^{2}}{Mr\beta}}q-3\sqrt2
e^{-2\sqrt2\sqrt\frac{q^{2}}{Mr\beta}}q\right),\quad
C=\left(1+e^{\sqrt2\sqrt\frac{q^{2}}{Mr\beta}}\right)^{2}Mr^{2}\beta,\\\label{4.5}
&D=-\frac{\sqrt2e^{-2\sqrt2\sqrt\frac{q^{2}}{Mr\beta}}q}{
\left(1+e^{-\sqrt2\sqrt\frac{q^{2}}{Mr\beta}}\right)^{3}Mr\beta\sqrt\frac{q^{2}}{Mr\beta}}.
\end{align}
Finally, we construct the Table \textbf{2} of numerical solutions of
phase transition from Eq.(\ref{4.6})
\begin{table}[ht]
\caption{Numerical solution of phase transition}
\centering 
\begin{tabular}{c c c c} 
\hline\hline 
cases & q=2 & q=5 & q=11 \\ [0.5ex] 
\hline 
T & 0.00041 & 0.00049 & 0.00057 \\ 
P & 0.9124 & 1.2410 & 2.2714 \\
M & 15.872 & 24.514 & 18.427 \\
S & 70.456 & 101.341 & 314.898 \\
r & 1.2711 & 3.2307 & 5.3217 \\
G & 0.6801 & 0.2502 & 0.2029 \\
$\zeta$ &0.0025 & 0.0075 & -0.0043 \\ [1ex ] 
\hline 
\end{tabular}
\label{table:critical points} 
\end{table}
\begin{figure}\center
\epsfig{file=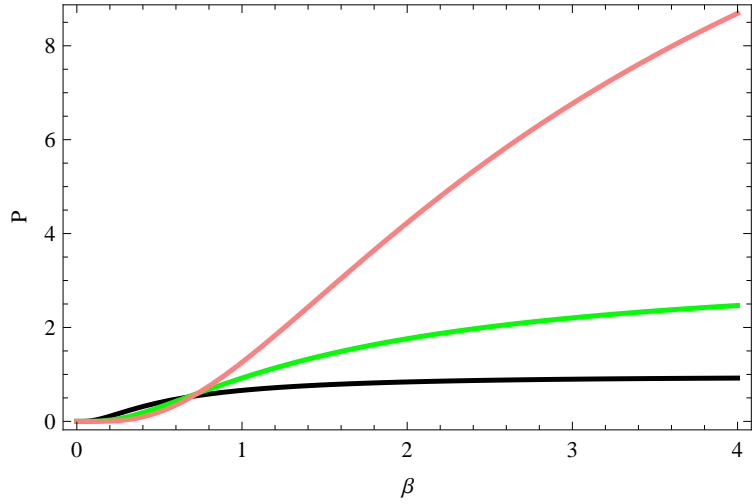,width=0.6\linewidth}\caption{$P-\nu$ curve corresponding to temperature for $q=2,q=5,q=11$.}
\end{figure}
\begin{figure}\center
\epsfig{file=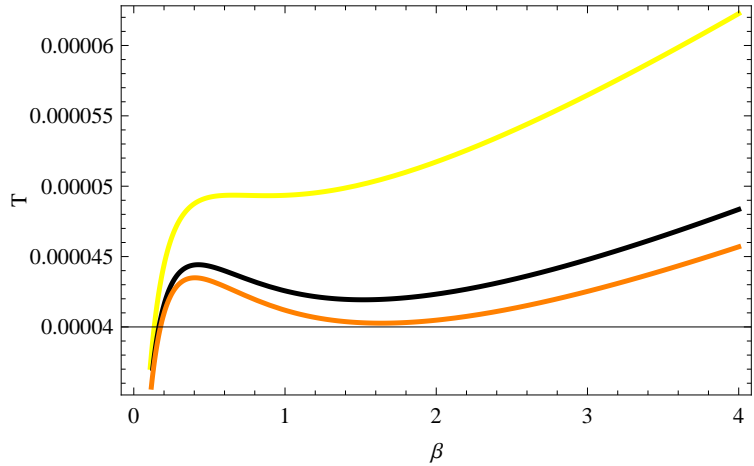,width=0.6\linewidth}\caption{$T-\beta$ curves showing consequence of volume expansitivity.}
\end{figure}
\begin{align}\label{4.6}
&T=\frac{2A\beta}{\pi},\\\nonumber
&P=\left[-2^{1+2\beta}\left(\frac{e^{-\sqrt2\sqrt\frac{q^{2}}{Mr\beta}}}
{(1+e^{-\sqrt2\sqrt\frac{q^{2}}{Mr\beta}})^{2}}\right)^{\beta-1}Mr
\left\{\frac{\sqrt2e^{-\sqrt2\sqrt\frac{q^{2}}{Mr\beta}}}q^{3}{\left(1+
e^{-\sqrt2\sqrt\frac{q^{2}}{Mr\beta}}\right)^{3}}r^{3}\right.\right.\\\nonumber
&\left.\left.\times \left(\frac{q^{2}}{Mr\beta}\right)^{\frac{3}{2}}M^{2}\beta^{2}
-\frac{B\left(\frac{q^{2}}{Mr\beta}\right)^{-1/2}}{C\left(1+e^{-\sqrt2\sqrt\frac{q^{2}}{Mr\beta}}\right)^{2}
Mr^{2}\beta}-\frac{e^{-2\sqrt2\sqrt\frac{q^{2}}{Mr\beta}}q^{3}}
{\sqrt2\left(1+e^{-\sqrt2\sqrt\frac{q^{2}}{Mr\beta}}\right)^{2}
M^{2}r^{3}\beta^{2}}\right.\right.\\\nonumber&\left.\left.\times\left(\frac{q^{2}}{Mr\beta}\right)^
{\frac{3}{2}}\right\}\beta
+2^{1+2\beta}\left(\frac{e^{-\sqrt2\sqrt\frac{q^{2}}{Mr\beta}}}{(1+
e^{-\sqrt2\sqrt\frac{q^{2}}{Mr\beta}})^{2}}\right)^{\beta-1}M(D)\beta-T2^{1+2\beta}
\left\{\left(e^{-\sqrt2\sqrt\frac{q^{2}}{Mr\beta}}\right)\right.\right.\\\label{4.7}
&\times \left.\left.\left(1+
e^{-\sqrt2\sqrt\frac{q^{2}}{Mr\beta}}\right)^{-2}\right\}^{\beta-2}Mr
\left(\frac{e^{-\sqrt2\sqrt\frac{q^{2}}{Mr\beta}}q^{2}(\frac{q^{2}}{Mr\beta})^{-1}}{\sqrt2\left(1+
e^{-\sqrt2\sqrt\frac{q^{2}}{Mr\beta}}\right)^{2}Mr^{2}\beta}-D
\right)(D)(\beta-1)\beta\right] ^{-1}.
\end{align}
Equation (\ref{3.7}) provides the following formulation in order to determine the location
of critical points,
\begin{equation}\label{4.8}
\left(\frac{\partial P}{\partial v_{\beta}}\right)_{T}=0,\quad
\left(\frac{\partial^{2} P}{\partial v^{2}_{\beta} }\right)_{T}=0.
\end{equation}
Equations (\ref{3.10})-(\ref{3.12}) gives Prigogine-Defay ratio $\prod$ to be $=2.13$ for the charged ABGB
using logistic distribution. It is well known that a phase transition is said to be glassy if $\prod$
ration is between $2-5$ for different cases of $q$.
Robert Tournier F described the concept of thermodynamics origin of the special transitions in his work \cite{Z18} as glassy
transition is require more modification but in our case we can only discuss PD ratio because here we have not require non-equilibrium thermodynamics quantities.

Moreover, for glassy phase transition, we have to conclude that
derivative of free energy is discontinuously behave as well as range of PD ratio, in all these cases glassy phase transition is tested by
PD ratio as Banerjee \emph{et al.} used in their work \cite{Z20}. Thus, we observed glassy phase transition in the charged ABGB BHs. Figures 8-11 describe the variations of some physical quantities with respect to few parametric values.

\section{Outlook}

In this paper, we have investigated some thermodynamical features of the static charged exact solutions of
the regular spherically relativistic structures. We have developed a systematic strategy to
shed light over the interesting phenomenon of phase transition in two viable charged models of BHs. First BH model
is supplemented with exponential distribution function and other is with logistic configuration variables. The
thermodynamical phenomenon of phase transition has been investigated for both of these models by divergences in
expansivity in stellar volume, specific heat and extent of compressibility near some important particular points.
After adopting grand canonical ensemble, we have carried our quantitative as well as qualitative investigation of
phase transition.

We would like to delineate the work of the thermodynamics and phase
transition of higher-dimensional regular ABGB BH (under various
distributions) with inner and outer horizons. We analyzed the phase
transition of ABGB BH,s shows similar behavior to van der Waals, value of $\upsilon$
showing changes at stability of system when
$\left(\frac{\partial P}{\partial v_{\beta}}\right)_{T}  < 0$ at
phase transition points with the help of specific values of electric charge q  where( $q=2,5,11$) and in opposite case
$\left(\frac{\partial P}{\partial v_{\beta}}\right)_{T} >0$ system will be unstable
. We ascertain
the critical points using precise curves, more than
that we interrogate the first and second order phase transition by
Ehrenfest's equation. In Section \textbf{3} we inspect the critical
behavior and phase transition of (ABGB)(using Distribution
function)according to beyond curves continuous deportment of Gibbs
free energy,entropy,critical effective  pressure,critical effective temperature and
volume expansivity coefficient. In Section.4 we remark the glassy phase
transition ,essentially these types of phase transition depends on specific
range of PD ratio.These outcomes execute the essence of liquid-gas
phase transition of ABGB at critical points therefor the
understanding of ABGB spacetime and its phase transition is very
influence.

In this paper, we summarize,dilate and determine in recent expansion
of phase transition in charged regular black hole
Ay\'{o}n-Beato-Garc\'{i}a-Bronnikov (ABGB) and contemplating the correlation
between the horizons, we consecrate the thermodynamics quantities
(using specific distribution) in $n=5$ dimensions. Furthermore, we
acquired critical effective temperature, pressure and volume expansivity $\beta$, also we
deliberate the critical behavior of specific regular black holes
which has lot of circumspection in recent years corresponding to the
rotating black hole, and conduct new phase transitions (glassy) for
specific range of PD ration $(\prod)$, besides thoroughly study of
Gibbs free energy, entropy and its behavior (under specified
distribution) by Ehrenfest equation. We look over the second order
phase transition at critical points. These results are compatible
with the essence the liquid-gas phase transition at the critical
points corresponding to work of Li, \emph{et. al.} \cite{Z2}.

\end{document}